\documentclass[5p]{elsarticle}

\usepackage{graphicx}
\usepackage{epstopdf}
\usepackage{epsfig}
\usepackage[english]{babel}
\usepackage{amsmath}
\usepackage{amssymb}

\usepackage{lineno,hyperref}
\modulolinenumbers[5]

\journal{Journal of Physica E}

\biboptions{sort&compress}
\begin{document}

\begin{frontmatter}

%
%%
%%\def\bsigma{\mbox{\boldmath $\sigma$}}
%%\def\bomega{\mbox{\boldmath $\omega$}}
%%\def\ss{\scriptscriptstyle }
%%%\def\baselinestretch{1.4}
%
%%\documentclass[aps,preprint,prl,showpacs,floatfix,superscriptaddress]{revtex4}
%%\documentclass[prb]{revtex4}
%%\documentclass[aps,preprint,prb,showpacs,floatfix,superscriptaddress]{revtex4}
%
%
%%\documentclass[aps,twocolumn,prb,showpacs,floatfix,superscriptaddress]{revtex4}
%%%%%%%%%%%%%%%%%%%%%%%%%%%%%%%%%%%%%%%%%%%%%%%%%%%%%%%%%%%%%%%%%%%%%%%%%%%%%%%%%%%%%%%%%%%%%%%%%%%%%%%%%%%%%%%%%%%%%%%%%%%%%%%%%%%%%%%%%%%%%%%%%%%%%%%%%%%%%%%%%%%%%%%%%%%%%%%%%%%%%%%%%%%%%%%%%%%%%%%%%%%%%%%%%%%%%%%%%%%%%%%%%%%%%%%%%%%%%%%%%%%%%%%%%%%%%
%%\usepackage{amsfonts}
%%\usepackage{graphicx}
%%\usepackage{amsmath}
%%\usepackage{amssymb}
%
%%\setcounter{MaxMatrixCols}{10}
%%TCIDATA{OutputFilter=LATEX.DLL}
%%TCIDATA{Version=5.00.0.2552}
%%TCIDATA{<META NAME="SaveForMode" CONTENT="1">}
%%TCIDATA{LastRevised=Tuesday, January 09, 2007 13:35:08}
%%TCIDATA{<META NAME="GraphicsSave" CONTENT="32">}
%%TCIDATA{Language=American English}
%
%%\input{tcilatex}
%
%%\begin{document}    in a quantum wire

\title{Self-consistent one-dimensional electron system  on liquid helium suspended over a nanoscale dielectric substrate}
\author[ufam]{Oleg G. Balev\corref{cor1}}
\ead{ogbalev@ufam.edu.br}

\author[ufca]{Antonio C.A. Ramos}
\ead{antonio.ramos@ufca.edu.br}

\cortext[cor1]{Corresponding author}

\address[ufam]{Departamento de F\'{\i}sica, Universidade Federal do Amazonas, 69077-000,
Manaus, Amazonas, Brazil}

\address[ufca]{Grupo de F\'{\i}sica Te\'orica e Computacional, Universidade Federal do Cariri, 63048-080, Juazeiro do Norte, Cear\'a, Brazil}

%\author{A.C.A. Ramos}
%\address{Grupo de F\'{\i}sica Te\'orica e Computacional, Universidade Federal do Cariri, 63048-080, Juazeiro do Norte, Cear\'a, Brazil}
%\ead{antonio.ramos@ufca.edu.br}

\begin{abstract} 
For electrons above a superfluid helium  film suspended on a specially designed dielectric substrate, $z=h(y)$, we obtain
that both the transverse, along $z$,  and  the lateral, along $y$,  quantizations are strongly enhanced due to  
a strong mutual coupling. 
The self-consistent quantum wires (QWs) with non-degenerated  one-dimensional electron systems (1DESs) are obtained 
over a superfluid liquid helium  (LH) suspended self-consistently  on different dielectric substrates with a nanoscale modulation.  
 A  gap $\gtrsim 10$meV  ($\gtrsim 1$meV) is obtained between the lowest two electron levels due to mainly the transverse (lateral)  quantization.
Our analytical model takes into account a strong interplay between the transverse and the lateral quantizations of an electron.
It uses that the characteristic length (energy) along the former direction is essentially
smaller (larger) than the one along the latter, in a close analogy with the adiabatic approximation. 
\end{abstract}


\end{frontmatter}

%\linenumbers %******************

%%%\date{\today}
%%%\pacs{73.21.-b, 75.75.-c,  73.20.Mf, 73.43.Lp}
%%%\maketitle
%67.25.bh   - Films and restricted geometries  /He^{4}/
%67.25.D    - Superfluid phase
%68.15.+e   - Liquid thin films
%   73.22.-f	Electronic structure of nanoscale materials and related systems
% 03.67.Lx	Quantum computation architectures and implementations
%73.21.-b  Electronic states and collective excitations in multilayers, quantum wells, mesoscopic, and nanoscale systems
%75.75.-c   Magnetic properties of nanostructures
%73.20.Mf   Cpllective excitations (including excitons, polarons, plasmons and other charge-density excitations)
%73.43.Lp   Edge Magnetoplasmons
%73.43.-f     Quantum Hall effects
%\maketitle

\section{Introduction}

Since pioneering works \cite{sommer1964,cole1969,shikin1970} quantized states of electrons above LH  %\textit{superfluid liquid helium} (LH), 
suspended on different substrates are the subject of a strong ongoing interest 
\cite{andrei1997,kovdrya2003,monarkha2004,ginzburg1978,monarkha1986,valkering1998,dykman1999,dahm2002,dykman2003,balev2004,
balev2005,lyon2006,lyon2008,ramos2008,dykman2010,lyon2011,kono2012,petrin2013,ramos2013,kono2015}. 
Electrons floating on LH have been proposed for quantum computing in a seminal work Ref. \cite{dykman1999}.
For a plane substrate and a large thickness of LH film, $d  \gtrsim 0.5 \mu$m, 
any effect of the substrate is negligible \cite{andrei1997,monarkha2004}. 
This allows a two-dimensional  electron system (2DES) on a bulk LH 
and  a single electron on a bulk film \cite{dykman1999} with a 1D hydrogenic 
spectrum \cite{andrei1997,kovdrya2003,monarkha2004,dykman1999}
$E_{m}^{1H}= - R/m^{2}$. Here $R \approx 8$K is an effective Rydberg energy. 
For quantum computing in Ref. \cite{dykman1999} 
it is suggested to patten the bottom electrode with features spaced 
close to $d$ ($ \approx 0.5 \mu$m). So that each feature traps one electron. 
Metallic posts submerged by the depth $\sim 0.5 \mu$m, beneath 
practically plane helium surface,  are suggested \cite{dykman2003}.  They form quantum dots 
for electrons on LH which may serve as the qubits of a quantum computer \cite{dykman2003}, in particular,
at temperature $T \approx 10$mK \cite{dykman1999,dykman2003}.  
Surface electrons with band-type spectrum on LH over metallic periodic substrate of the 
diffraction grating type are proposed by Ginzburg and Monarkha \cite{ginzburg1978}.
Where an amplitude of modulation is much smaller than $d$ and a free surface of LH
is assumed as flat. 

Electrons in a micron-scale and a nanoscale channels filled by capillary action with
LH  \cite{kovdrya2003,monarkha1986,valkering1998,dykman1999,balev2004,balev2005,lyon2006,lyon2008,ramos2008,dykman2010,lyon2011,kono2012,
petrin2013,ramos2013,kono2015} attract recently much attention, in particular, due to their high potential in creating
qubits with the needed properties of performance.   
The systems of such channels are promising for construction of the equivalent of a charge-coupled device (CCD) \cite{boyle1970} that, in addition, 
will allow the large scale transport of qubits \cite{lyon2006,lyon2008,lyon2011}. In interesting experiments of Refs. \cite{lyon2008,lyon2011}
electrons are studied in the channels of a width $\gtrsim 3 \mu$m at $T \approx 1.5$K.
Theoretical framework of Refs.  \cite{lyon2008,lyon2011} treats electrons mainly as ones above  a bulk LH. 
%%%************

\begin{figure}[ht]
\vspace*{-0.0cm}%-0.3%0.5%-1.
\includegraphics [height=4cm, width=8.5cm]{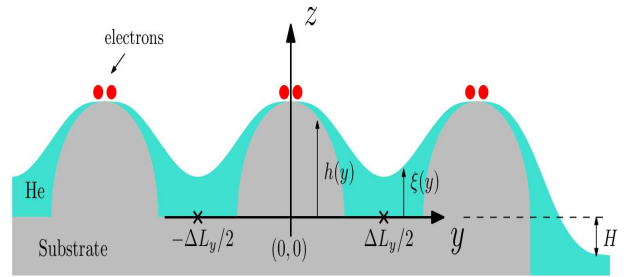}
\vspace*{-0.2cm}%-1.0%-3.0%-4.5cm%-2.0%-0.7
\caption{(Color online) A sketch, not to scale, of a model geometry. Substrate surface $z=h(y)$, LH surface $z=\xi(y)$, and LH thickness $d(y)=\xi(y)-h(y)$.
}
\label{fig1}
\end{figure}

Indeed, usually some important characteristics of electrons in these
channels such as a gap between the lowest electron levels of the transverse (lateral) quantization, 
a form of the lateral potential, a form of the LH surface, a lateral density profile, etc. are not well known
\cite{kovdrya2003,monarkha1986,valkering1998,balev2005,lyon2006,lyon2008,ramos2008,lyon2011,kono2012,
petrin2013,ramos2013,kono2015}.
In particular, due to absence  of interplay and self-consistency between transverse and lateral quantizations  within used theoretical frameworks.

In present study  self-consistent 1DESs in QWs over LH suspended on different nanoscale dielectric substrates are obtained, for  $T=0.6$K.  
A strong interplay between the quantizations of an electron along $z$ and $y$ directions  
is treated within present approach. It uses, in particular,  some analogies with 
well known adiabatic approximation \cite{anselm1978}. In Fig. \ref{fig1}  the sketch of a geometry of studied model is shown. 
We consider that a dielectric substrate is periodic along the $y$-direction with a finite period $\Delta L_{y}$ (unless otherwise stated)
and $L_{x} \to \infty$. 
Within the main super cell  ($L_{x} \times \Delta L_{y}$; $|y| \leq \Delta L_{y}/2$) the substrate profile $z=h(y)$ is assumed as
% $z=h(y)$ is assumed as
\begin{eqnarray}
&& h(y)=0 \; , \;\;\; a/4 \leq |y| \leq \Delta L_{y}/2 \;\; , \nonumber \\
&& h(y)=h_{1} \cos(2\pi y/a) \; , \; \;\;  |y| \leq a/4  \; ,
 \label{eq1}
\end{eqnarray}
where $h_{1}$, $a$, $\Delta L_{y}$ are the characteristic  scales of its modulation, cf. Fig. \ref{fig1}. We assume that  $2h_{1}/a \ll 1$. I.e., 
a substrate profile is smooth.

For obtained systems of QWs, within the main super cell a 1DES is laterally localized at $y=0$, cf. Fig. \ref{fig1}.
Point out that effect of tunnel coupling between 1DESs of neighboring super cells is negligible 
for present systems of  QWs. 
We have obtained a strong  "long-range" effect of $\Delta L_{y}$ on the properties of a self-consistent 
1DES at the region $10 \mu$m$\geq \Delta L_{y} \geq 1 \mu$m. For a given linear density within a super cell $n_{L}=N_{tot}/L_{x}$;
$N_{tot}$ is the total number of electrons within a super cell.
It is related with an essential dependence of  LH profile within this region.
That induces a strong modification of the transverse and the lateral quantizations for an electron.
In particular, an essential modification of the effective electron potential
is obtained due to a strong change for the image potential of substrate.
 
 Notice, for $\Delta L_{y} \gtrsim 50 \mu$m properties of a self-consistent 
1DES become practically independent of $\Delta L_{y}$. In present figures
we assume that $\Delta L_{y}=1 \mu$m or $10 \mu$m. Then obtained results can 
be applied to the properties of QWs within a finite region $|y| \leq L_{y}/2$, 
with $L_{y }\geq \Delta L_{y}$, if the substrate have a finite region of periodic 
modulation $|y| \geq L_{y}/2+25 \mu$m.

In Subsection 2.1  we present a self-consistent Hamiltonian of an electron on a self-consistent LH film, suspended over
a dielectric substrate with a nanoscale lateral modulation.  In Subsection 2.2 we give the rest of a self-consistent 
framework for our model.  It defines a self-consistent profile of LH suspended on the dielectric substrate, for a given linear density within
a super cell. In Section 3 we present results and discussions on the self-consistent
profiles of LH films suspended on the special dielectric substrates, the lowest levels of the transverse and of the lateral
quantizations,  a self-consistent electron density $n(y)$ profiles of 1DESs in obtained self-consistent electron nano-channels. 
Conclusions follow in Section 4.

\section{Self-consistent model of electrons over liquid helium on a substrate with nanoscale modulation} 

\subsection{One-electron Hamiltonian} 

We consider that between a surface of LH, $z= \xi(y)$, and the surface of substrate,
$z=h(y)$, Eq. (\ref{eq1}) a LH film is formed  of the thickness $d(y)=\xi(y)-h(y)>0$.
First we assume that $\Delta L_{y} \to \infty$, later on we will show how our study can be extended to
a finite  $\Delta L_{y}$.
Then the wave functions and the eigenvalues of an electron over LH are defined  by the Schrodinger equation \cite{kovdrya2003,monarkha2004,balev2005} 
%Assuming that the surface of helium film is given by $z=z_{hs}(y) \equiv \xi(y)$ (i.e., its z-coordinate depends only on $y$),
%we have on the substrate, $z=z_{su}(y) \equiv h(y)$,  a helium film of the thickness $d(y)=\xi(y)-h(y)>0$.
%The wave functions and eigenvalues of an electron over the liquid helium (LH) are defined  by the following
%Schrodinger equation  

\begin{eqnarray}
&&\left[-\frac{\hbar^2}{2m_{0}}\left(\frac{\partial^{2}}{\partial z^{2}}+\frac{\partial^{2}}{\partial y^{2}} +\frac{\partial^{2}}{\partial x^{2}}  \right) +
V(z,y)\right] \nonumber \\
&&\times  \Psi_{\beta}(z,y,x) =W_{\beta} \Psi_{\beta}(z,y,x) ,
 \label{eq2}
\end{eqnarray}
where three quantum numbers $\beta=\{k_{x\beta},n_{y\beta},n_{z\beta}\}$ are given by the wave number $k_{x\beta}=2\pi n_{x\beta}/L_{x}$ and two integer quantum numbers $n_{y\beta}=1, 2, 3,...$, $n_{z\beta}=1, 2, 3,...$. As we assume the Born-von Karman boundary condition along $x$, 
we have $n_{x\beta}=0, \pm 1, \pm 2,...$. 
In Eq. (\ref{eq2}),  e.g., following Refs. \cite{kovdrya2003,monarkha2004,balev2005},  we have that 
\begin{equation}
V(z,y)= -\frac{\Lambda}{z-\xi(y)} -\frac{\Lambda_{1}}{z-h(y)} + |e| E_{p} z ,
 \label{eq3}
\end{equation}
where $\Lambda=e^2(\varepsilon_{LH}-1)/[4(\varepsilon_{LH}+1)]$ and $\Lambda_{1}=e^2(\varepsilon_{S}-1)/[4(\varepsilon_{S}+1)]$.
Here $\varepsilon_{LH} \approx 1.054$ 
%%%1.057(1.054???)$ 
is the dielectric constant of LH, $\varepsilon_{S}$ is the dielectric constant of substrate, and
$E_{p}$ is an external (also called as holding) electric field.  The first two terms in the  right hand side of Eq. (\ref{eq3}) represent
the main contributions to the image potential energy \cite{monarkha2004}. The former term represents the image potential energy due to a bulk
LH and the latter one shows a main effect of the substrate (for an infinite thickness of a LH film it is nullified).

Point out, Eq. (\ref{eq3}) can be considered as exact if $\xi(y)$ and $h(y)$ are the linear polynomial functions of $y$  or independent of $y$.
For more complex dependences of $\xi(y)$ and $h(y)$ on $y$, Eq. (\ref{eq3}) is valid  if $h(y)$  is smooth enough within an actual
region. Where an electron is present mainly. This justifies the second term in Eq. (\ref{eq3}).
Further, the first term in Eq. (\ref{eq3}) is readily justified due to a smoother $\xi(y)$ than $h(y)$ and closer average position of an electron along $z$ to
the characteristic boundary. Here it is the LH surface $\xi(y)$. 
I.e., in Eq.(3) an electron image potential is well approximated by the first two terms of the right hand side 
provided the distance between the electron and the dielectric is small relative to the curvature 
of the dielectric surface.

As potential Eq. (\ref{eq3}) is independent of $x$ we look for a solution of Eq. (\ref{eq2}) as follows
\begin{equation}
\Psi_{\beta}(z,y,x) = L_{x}^{-1/2} e^{ik_{x\beta} x} \; \psi_{n_{z\beta}, \;n_{y\beta}}(z,y) \;   .
 \label{eq4}
\end{equation}
Then from Eq. (\ref{eq2}) we obtain 
\begin{eqnarray}
&&\left[-\frac{\hbar^2}{2m_{0}}\left(\frac{\partial^{2}}{\partial z^{2}}+\frac{\partial^{2}}{\partial y^{2}}  \right) +V(z,y)\right] 
\psi_{n_{z\beta}, \;n_{y\beta}}(z,y) \nonumber \\
&& =
\widetilde{W}_{n_{z\beta}, \;n_{y\beta}} \psi_{n_{z\beta}, \;n_{y\beta}}(z,y) ,
 \label{eq5}
\end{eqnarray}
where $\widetilde{W}_{n_{z\beta}, \;n_{y\beta}}=W_{\beta}-\frac{\hbar^{2} k_{x\beta}^{2}}{2m_{0}} $.

\begin{figure}[ht]
\vspace*{-0.7cm}%-0.3   %0.5%-1.5
\includegraphics [height=13.0cm, width=10cm]{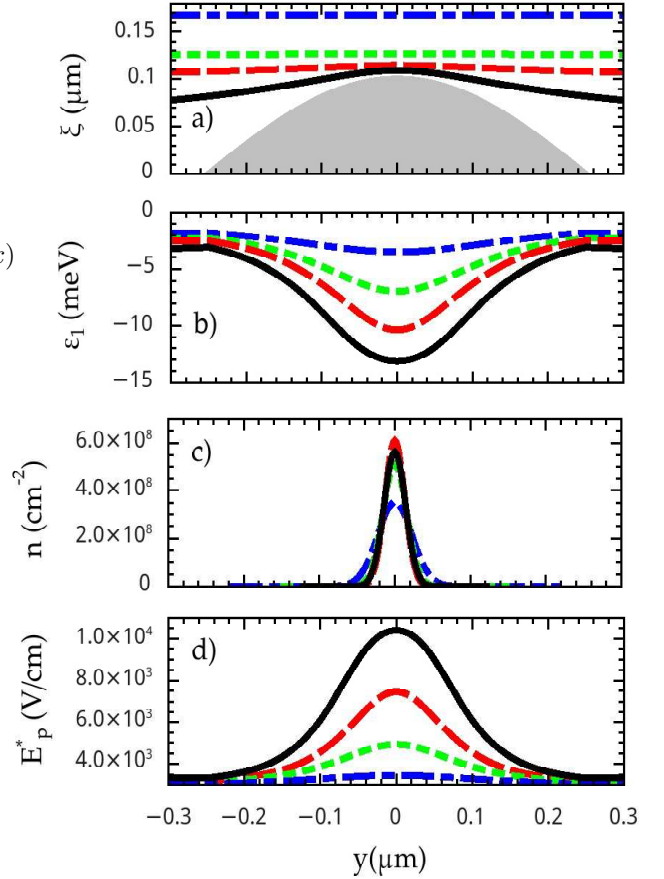}
\vspace*{-0.7cm}%-0.7;-1.0%-3.0%-4.5cm%-2.0%-0.7
\caption{(Color online) By the solid curves, for $H=10$cm, the dashed curves, for $H=2.5$cm,  
 the dotted curves, for $H=0.5$cm, and the dot-dashed curves, for $H=0.05$cm,  
are plotted on panel: 
 (a) the spatial profile of LH surface $\xi$ (the substrate is shown by a grey background color);
(b) the spatial dependence of the ground state energy $\mathcal{E}_{1}$, due to transverse quantization;
(c) the spatial dependence of the surface electron density $n$; 
(d) the spatial dependence of the effective electric field
$E_{p}^{\ast}$ on an electron, in the ground state $\mathcal{E}_{1}$.
 Here we have $h_{1}=0.1 \mu$m, $a=1.0 \mu$m, $\Delta L_{y}=1 \mu$m.  
 For Figs. \ref{fig2} - \ref{fig11} $T=0.6$K, $\varepsilon_{S}=5$, $n_{L}=2 \times 10^{3} cm^{-1}$,  and $E_{p}=5 V/cm$
 are common. $z=-H$ is the level of bulk LH.} 
\label{fig2}
\end{figure}

\begin{figure}[ht]
\vspace*{-0.9cm}%-0.3%0.5%-1.5
\includegraphics [height=9cm, width=10.5cm]{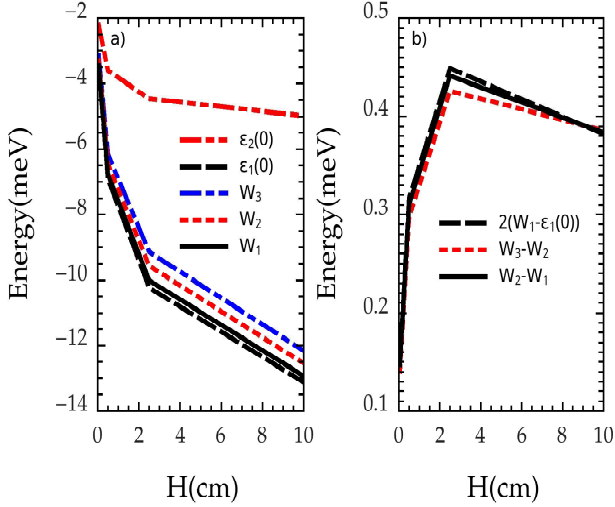}
\vspace*{-0.7cm}%-1.0%-3.0%-4.5cm%-2.0%-0.7
\caption{(Color online) The energy spectra are shown as functions of $H$ for
conditions of Fig. \ref{fig2}.
%, at $10$ cm$\geq H \geq 0.05\;$ cm. 
On panel (a) the solid, the dotted, and the dash-dot curves plot three the lowest levels of the lateral 
quantization $W_{1}, W_{2}$, and $W_{3}$. The dashed curve and the
dot-dot-dash curve plot  $\mathcal{E}_{1}(0)$ and  $\mathcal{E}_{2}(0)$, i.e., two the lowest 
levels of the transverse quantization at $y=0$. On panel
(b) the solid, the dashed, and  the dotted curves plot $(W_{2}-W_{1})$, $2(W_{1}-\mathcal{E}_{1}(0))$,
and $(W_{3}-W_{2})$. As these three curves are very close, in an actual region
the effective lateral potential $\mathcal{E}_{1}(y)$ for 1DES is very close to a parabolic one.} 
\label{fig3}
\end{figure}

\begin{figure}[ht]
\vspace*{-0.3cm}%-0.3%0.5%-1.5
\includegraphics [height=14cm, width=10cm]{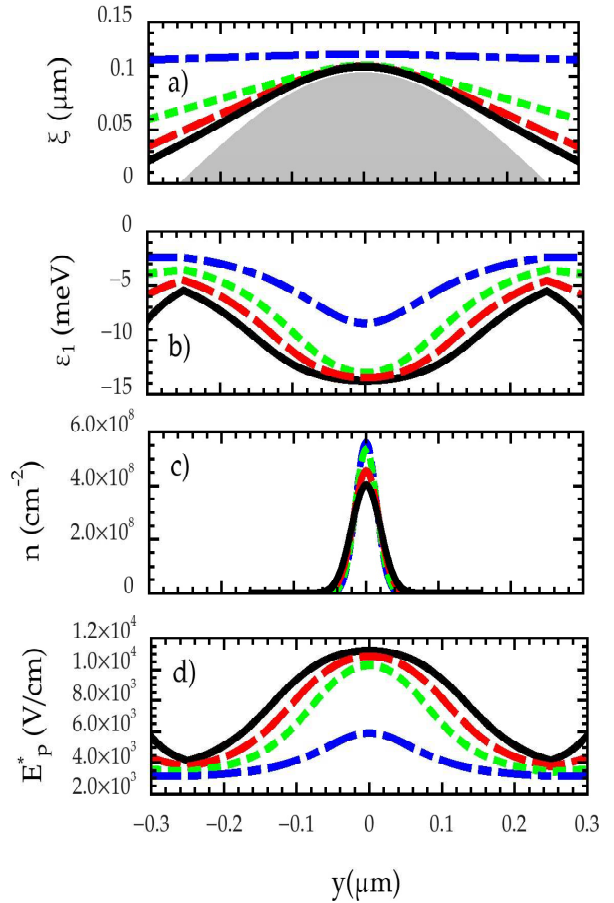}
\vspace*{-0.7cm}%-1.0%-3.0%-4.5cm%-2.0%-0.7
\caption{(Color online)  Same dependences as in Fig. \ref{fig2}, for the same conditions except $\Delta L_{y}=10\mu$m.} 
\label{fig4}
\end{figure}

\begin{figure}[ht]
\vspace*{-0.9cm}%-0.3%0.5%-1.5
\includegraphics [height=9cm, width=10.5cm]{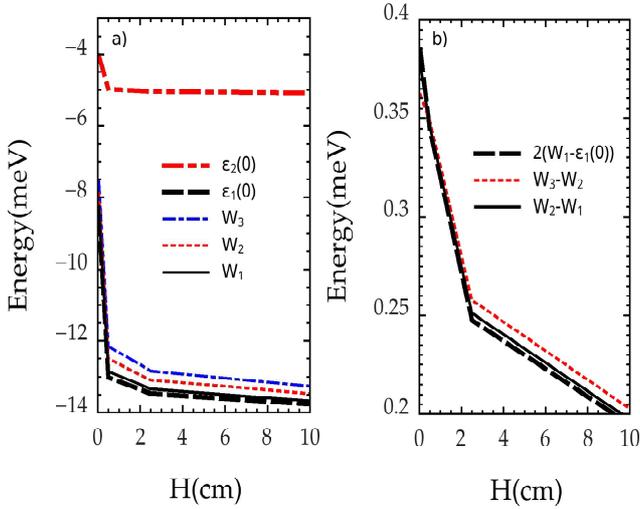}
\vspace*{-0.7cm}%-1.0%-3.0%-4.5cm%-2.0%-0.7
\caption{(Color online) The same energy spectra as in Fig. \ref{fig3} for conditions of Fig. \ref{fig4}. 
Three curves on panel (b) are very close, same as in Fig. \ref{fig3}(b). It shows that in an actual region
the effective lateral potential $\mathcal{E}_{1}(y)$ for 1DES is very close to a parabolic one.}
\label{fig5}
\end{figure}

\begin{figure}[ht]
\vspace*{-0.3cm}%-0.3%0.5%-1.5
\includegraphics [height=14cm, width=10cm]{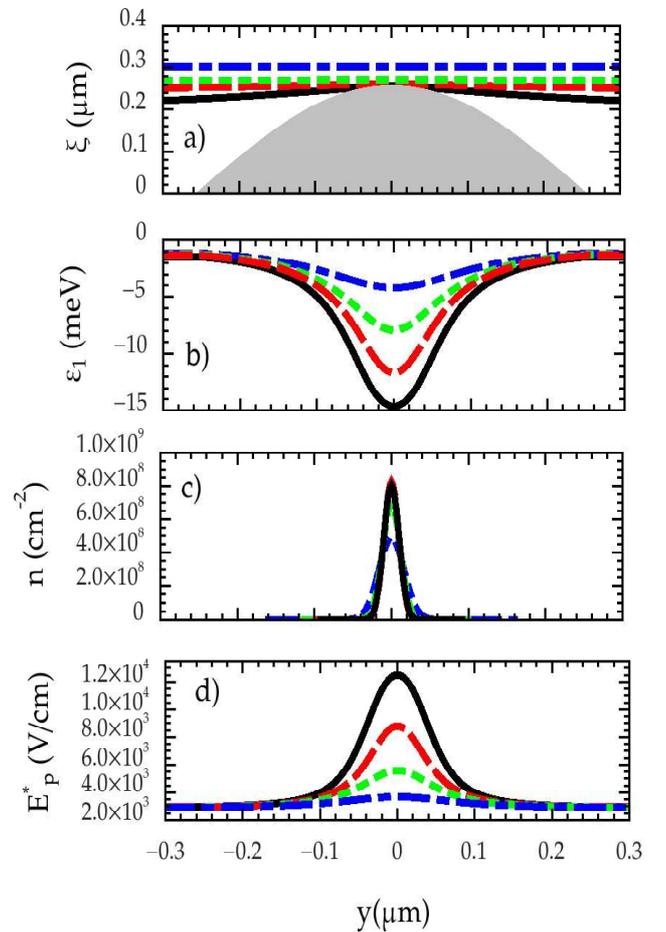}
\vspace*{-0.7cm}%-1.0%-3.0%-4.5cm%-2.0%-0.7
\caption{(Color online) Same dependences as in Fig. \ref{fig2}, for the same conditions apart from $h_{1}=0.25 \mu$m;
notice, here $\Delta L_{y}=1 \mu$m.}
\label{fig6}
\end{figure}

\begin{figure}[ht]
\vspace*{-0.9cm}%-0.3%0.5%-1.5
\includegraphics [height=9cm, width=10.5cm]{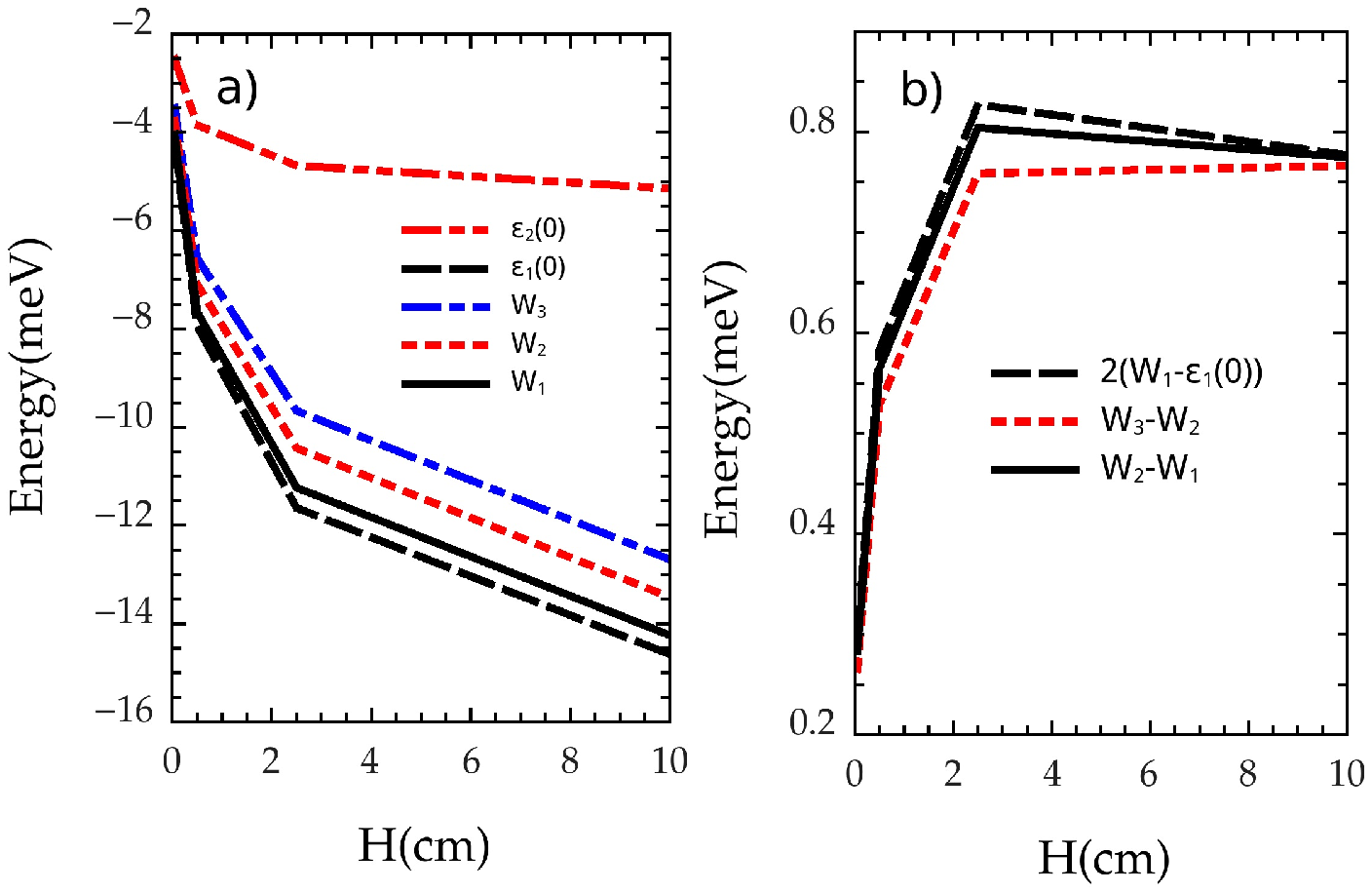}
\vspace*{-0.7cm}%-1.0%-3.0%-4.5cm%-2.0%-0.7
\caption{(Color online) The same energy spectra as in Fig. \ref{fig3} for conditions of Fig. \ref{fig6}. 
Three curves on panel (b) are close, similar with Fig. \ref{fig3}(b). It shows that in an actual region
the effective lateral potential $\mathcal{E}_{1}(y)$ for 1DES is well approximated by a parabolic one.}
\label{fig7}
\end{figure}

\begin{figure}[ht]
\vspace*{-0.3cm}%-0.3%0.5%-1.5
\includegraphics [height=14cm, width=10cm]{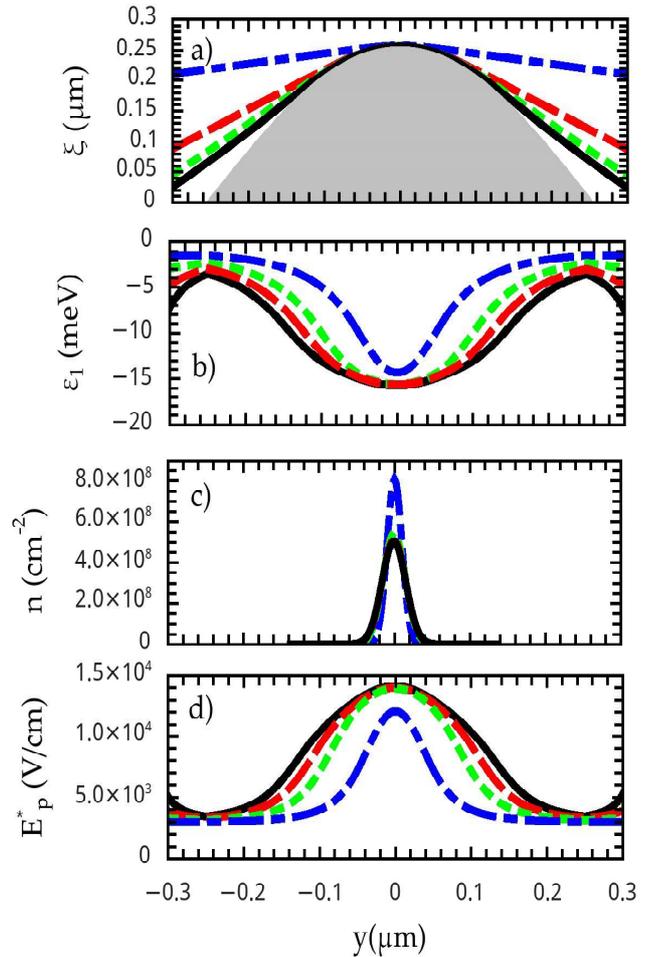}
\vspace*{-0.7cm}%-1.0%-3.0%-4.5cm%-2.0%-0.7
\caption{(Color online) Same dependences as in Fig. \ref{fig6}, for the same conditions apart from $\Delta L_{y}=10 \mu$m.}
\label{fig8}
\end{figure}

\begin{figure}[ht]
\vspace*{-0.9cm}%-0.3%0.5%-1.5
\includegraphics [height=9cm, width=10.5cm]{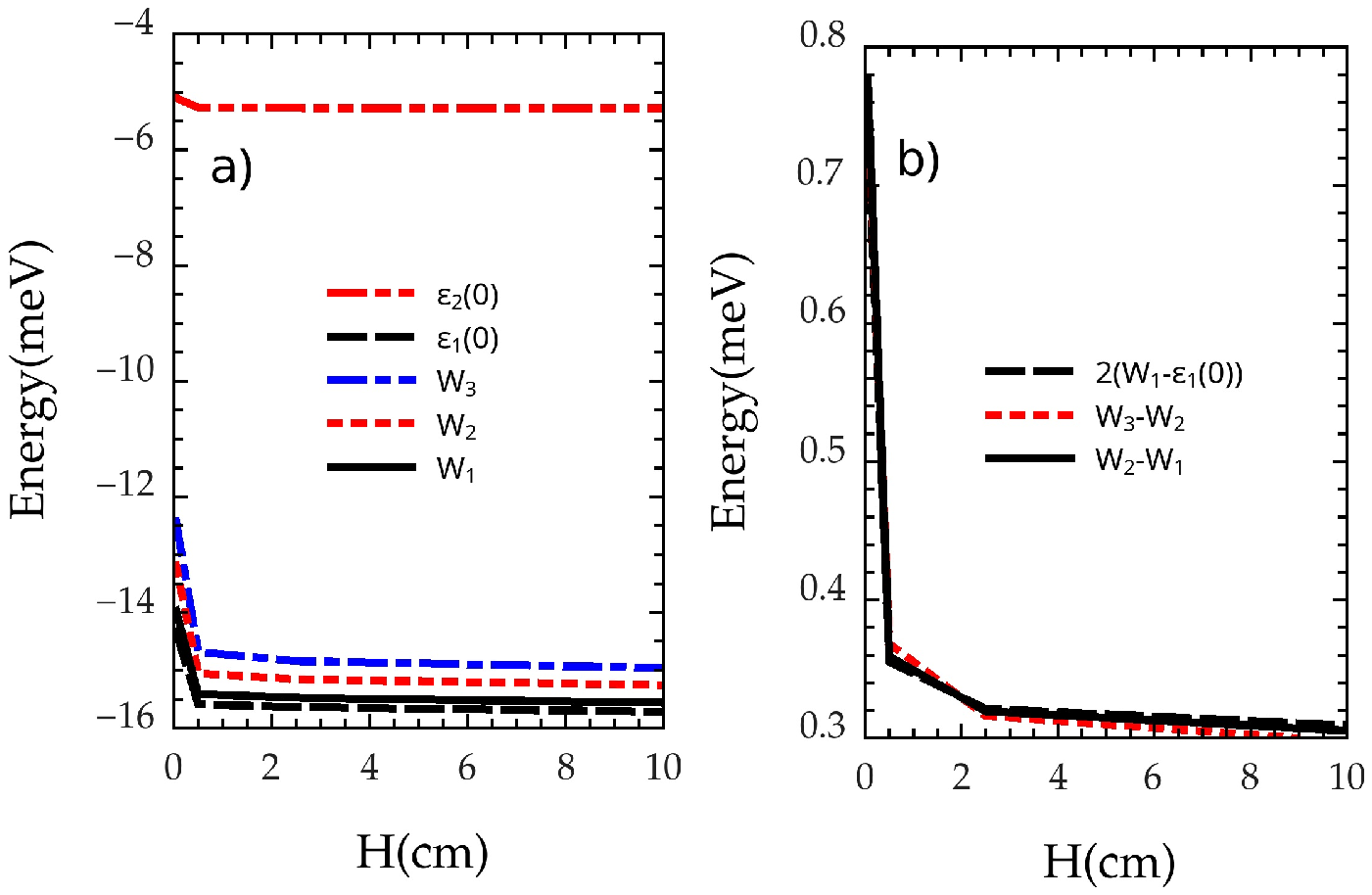}
\vspace*{-0.7cm}%-1.0%-3.0%-4.5cm%-2.0%-0.7
\caption{(Color online) The same energy spectra as in Fig. \ref{fig7} for conditions of Fig. \ref{fig8}. 
Three curves on panel (b) are close, similar with Fig. \ref{fig7}(b). This shows that in an actual region
the effective lateral potential $\mathcal{E}_{1}(y)$ for 1DES is very well approximated as a parabolic one.}
\label{fig9}
\end{figure}

\begin{figure}[ht]
\vspace*{-0.7cm}%-0.3%0.5%-1.5
\includegraphics [height=14cm, width=10cm]{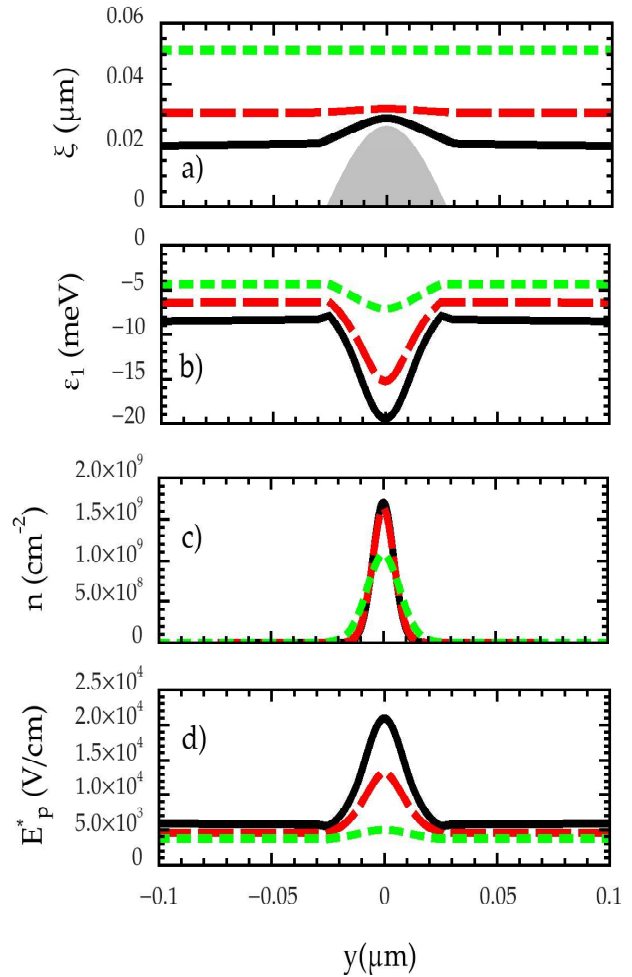}
\vspace*{-1.2cm}%-1.0%-3.0%-4.5cm%-2.0%-0.7
\caption{(Color online) Same dependences as in Fig. \ref{fig2} or Fig. \ref{fig6}, for the same conditions except for: 
$h_{1}=0.025 \mu$m, $a=0.1 \mu$m. In addition, the curves for $H=0.05$cm are absent.
% as here some conditions for a 1DES could not be satisfied. 
 }
\label{fig10}
\end{figure}

\begin{figure}[ht]
\vspace*{-0.9cm}%-0.3%0.5%-1.5
\includegraphics [height=9cm, width=10.5cm]{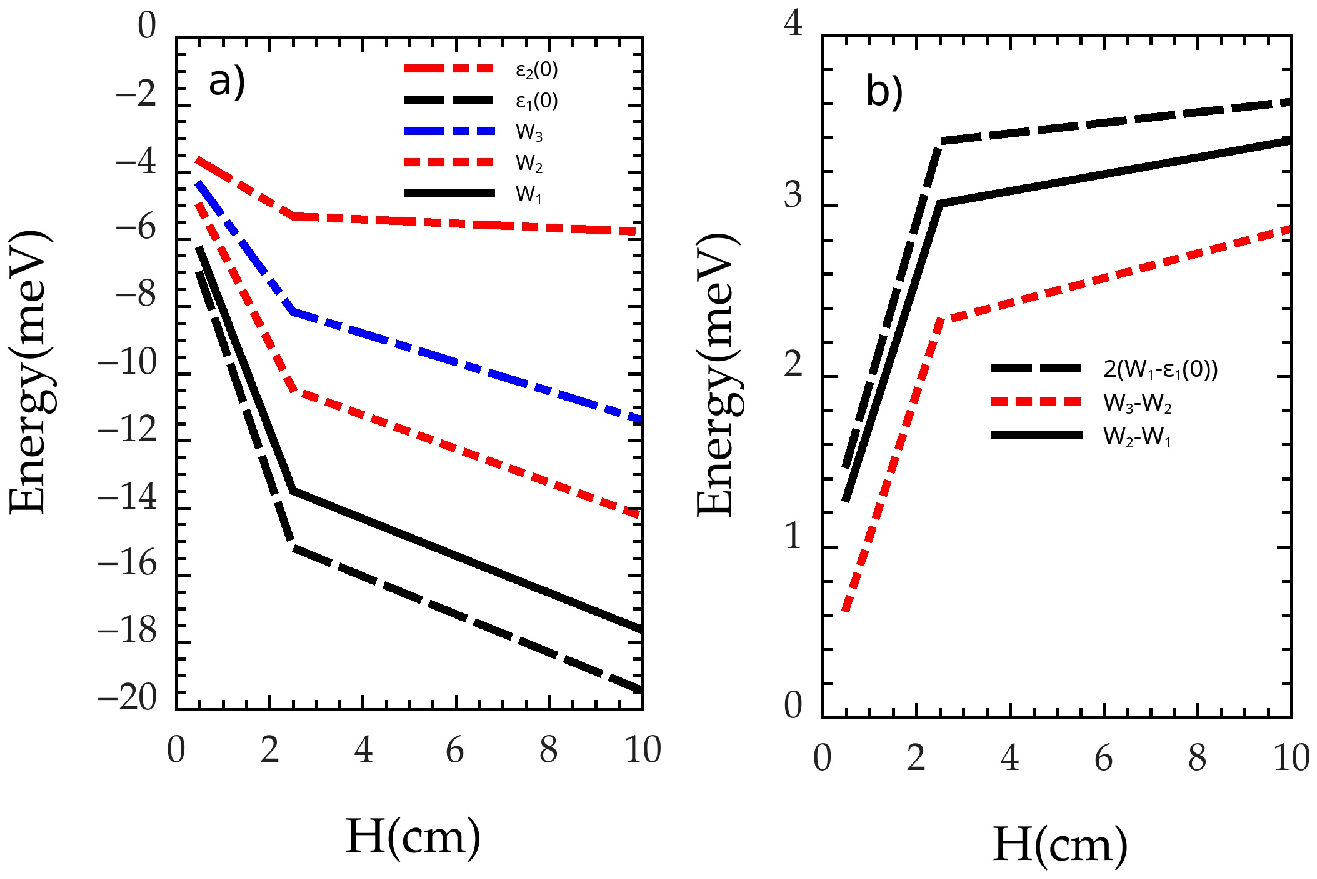}
\vspace*{-0.7cm}%-1.0%-3.0%-4.5cm%-2.0%-0.7
\caption{(Color online) The same energy spectra as in Fig. \ref{fig7} are plotted  for conditions of Fig. \ref{fig10}; here $10$ cm$\geq H \geq 0.5$ cm}
\label{fig11}
\end{figure}

To solve Eq. (\ref{eq5}) we develop an approach similar with the well known adiabatic method \cite{anselm1978},  that separates
a fast movement of electrons from a slow movement of nuclei, to separate a fast movement along z-axis, on a short  space scale $\Delta z$,
 from a slow movement along y-axis, on the scale $\Delta y \gg \Delta z$.  We assume that
\begin{equation}
\psi_{n_{z\beta}, \;n_{y\beta}}(z,y)= \Phi_{n_{y\beta}}(y) \varphi_{n_{z\beta}}(z,y) ,
 \label{eq6}
\end{equation}
where $\varphi_{n_{z\beta}}(z,y)$ is a real function (this condition always can be satisfied as it is a discrete spectrum state;
$n_{z\beta}=1, 2,...$) that satisfies
\begin{equation}
\left[-\frac{\hbar^2}{2m_{0}} \frac{\partial^{2}}{\partial z^{2}}  +V(z,y) \right] \varphi_{n_{z\beta}}(z,y)=
\mathcal{E}_{n_{z\beta}}(y) \varphi_{n_{z\beta}}(z,y) ,
 \label{eq7}
\end{equation}
where $y$ has the role of a parameter.
% in Eq. (\ref{eq7}). 
Then, substituting Eq. (\ref{eq6}) in Eq. (\ref{eq5}) and using Eq. (\ref{eq7}), we obtain
\begin{eqnarray}
&&-\frac{\hbar^2}{2m_{0}} \left[\varphi_{n_{z\beta}}(z,y)  \frac{\partial^{2}}{\partial y^{2}} \Phi_{n_{y\beta}}(y) + 
2 \frac{\partial \Phi_{n_{y\beta}}(y)}{\partial y}   \right. \nonumber \\
&& \left. \times \frac{\partial \varphi_{n_{z\beta}}(z,y)}{\partial y} 
+\Phi_{n_{y\beta}}(y) \frac{\partial^{2}}{\partial y^{2}} \varphi_{n_{z\beta}}(z,y) \right]  \nonumber \\
&&= (\widetilde{W}_{n_{z\beta}, \;n_{y\beta}}-\mathcal{E}_{n_{z\beta}}(y)) \Phi_{n_{y\beta}}(y) \varphi_{n_{z\beta}}(z,y) .
 \label{eq8}
\end{eqnarray}
As a wave function of discrete spectrum $\varphi_{n_{z\beta}}(z,y)=0$, for $z \leq \xi(y)$, and it is localised at $z \approx \xi(y)$
(e.g., within a few nanometers from the LH surface for typical conditions of below Figs. \ref{fig2} - \ref{fig11}), we obtain from
its normalization
\begin{equation}
\int_{-\infty}^{\infty} dz \; \varphi^{2}_{n_{z\beta}}(z,y) =1 ,
 \label{eq9}
\end{equation}
after applying $\partial/ \partial y$, that
\begin{equation}
\int_{-\infty}^{\infty} dz \; \varphi_{n_{z\beta}}(z,y) \partial  \varphi_{n_{z\beta}}(z,y)/\partial y =0 .
 \label{eq10}
\end{equation}

Then multiplying Eq. (\ref{eq8}) by $\varphi_{n_{z\beta}}(z,y)$ and integrating over z, $\int_{-\infty}^{\infty} dz $,  and using Eqs. (\ref{eq9})-(\ref{eq10}),
we obtain
\begin{eqnarray}
&&-\frac{\hbar^2}{2m_{0}} \frac{d^{2}}{d y^{2}} \Phi_{n_{y\beta}}(y) +\left[ \mathcal{E}_{n_{z\beta}}(y)
-\frac{\hbar^2}{2m_{0}}  \right. \nonumber \\
&&  \times \int_{-\infty}^{\infty} dz \; \varphi_{n_{z\beta}}(z,y)  \left. \frac{\partial^{2}}{\partial y^{2}} \varphi_{n_{z\beta}}(z,y) \right] \Phi_{n_{y\beta}}(y)   \nonumber \\
&&= \widetilde{W}_{n_{z\beta}, \;n_{y\beta}} \Phi_{n_{y\beta}}(y)  .
 \label{eq11}
\end{eqnarray}
%Using Eq. (\ref{eq7}), we estimate that the second term (i.e., the integral one) in the square brackets of Eq. (\ref{eq11}) gives only a very small, of the order  of 
%$(\Lambda_{z}/\Lambda_{y})^2 \ll 1$, correction to the first term, $\mathcal{E}_{n_{z\beta}}(y)$, and can be neglected. Then,
%similar with the adiabatic approximation, we have
%%%
Using Eq. (\ref{eq7}) and $(\Delta z/ \Delta y)^2 \ll 1$, we assume that the second term in the square brackets 
of Eq. (\ref{eq11}) gives only a small correction to the first term, $\mathcal{E}_{n_{z\beta}}(y)$, and can be neglected.
More accurate condition of a smallness of the non-adiabatic term in Eq. (\ref{eq11}) is given below by Eq. (\ref{eq22}). In particular, for studied 
in Figs \ref{fig2} - \ref{fig11} self-consistent quantum wires  over LH 
%with nanoscale width of electron distribution 
non-adiabatic contributions in Eq. (\ref{eq11}) 
are very small. Then, similar with the adiabatic approximation, we have
\begin{eqnarray}
&&-\frac{\hbar^2}{2m_{0}} \frac{d^{2}}{d y^{2}} \Phi_{n_{y\beta}}(y;n_{z\beta}) +\mathcal{E}_{n_{z\beta}}(y) \Phi_{n_{y\beta}}(y;n_{z\beta})  
\nonumber \\
&&= \widetilde{W}_{n_{z\beta}, \;n_{y\beta}} \Phi_{n_{y\beta}}(y;n_{z\beta})  ,
 \label{eq12}
\end{eqnarray}
where $n_{z\beta}$ is a discrete parameter;  usually we will be interested in the lowest level $n_{z\beta}=1$
and the first exited level $n_{z\beta}=2$ of Eq. (\ref{eq7}). Here we show explicitly
a dependence of the wave function $\Phi_{n_{y\beta}}(y)$ on $n_{z\beta}$ (cf. with Eqs. (\ref{eq6}), (\ref{eq8}), (\ref{eq11}) )
 as  $\Phi_{n_{y\beta}}(y;n_{z\beta}) $.
 Notice, it is  assumed that wave functions $\Phi_{n_{y\beta}}(y;n_{z\beta})$ are normalized within
the main super cell, $\int_{-\Delta L_{y}/2}^{\Delta L_{y}/2} dy \; |\Phi_{n_{y\beta}}(y;n_{z\beta})|^{2}=1$, for $\Delta L_{y} \to \infty$.
However,  present treatment also holds with a finite $\Delta L_{y}$ for the lowest levels $\widetilde{W}_{n_{z\beta}, \;n_{y\beta}}$
occupied by electrons if the tunnel coupling between such states in neighboring super cells is negligible for any realistic 
properties of experimental setup. In particular, as well for a very high quality setups with a very small effect of disorder on these energy levels.

Now we rewrite Eq. (\ref{eq7}) using, instead of $z$, a new variable $\tilde{z}=z-\xi(y)$ as
\begin{eqnarray}
&&\left[-\frac{\hbar^2}{2m_{0}} \frac{\partial^{2}}{\partial \tilde{z}^{2}}  +
\left(  -\frac{\Lambda}{\tilde{z}} -\frac{\Lambda_{1}}{\tilde{z}+d(y)} \right. \right. \nonumber \\
&& \left.  \left. + |e| E_{p} (\tilde{z}+\xi(y) )   \right)  \right]  \varphi_{n_{z\beta}}(\tilde{z}+\xi(y),y) \nonumber \\
&& = \mathcal{E}_{n_{z\beta}}(y) \varphi_{n_{z\beta}}(\tilde{z}+\xi(y),y)  .
 \label{eq13}
\end{eqnarray}
Let us introduce the characteristic scales of the length $a_{0}=\hbar^{2}/ \Lambda  m_{0} \approx 76 \; \AA$, of the time $t_{0}=\hbar^{3}/m_{0} \Lambda^{2}$,
of the energy $E_{0}=m_{0} \Lambda^{2}/\hbar^{2} \approx 16 \;$K, and a dimensionless variable $x=\tilde{z}/a_{0}$ . Then Eq. (\ref{eq13}) we
rewrite, with $\varphi^{y}_{n_{z\beta}}(x)=\varphi_{n_{z\beta}}(a_{0} x+\xi(y),y)$, as
\begin{eqnarray}
&&\left[ \frac{\partial^{2}}{\partial x^{2}}  + 2
\left( \frac{\mathcal{E}_{n_{z\beta}}(y)}{E_{0}}  +\frac{1}{x} +\frac{\Lambda_{1}/\Lambda}{x+d(y)/a_{0}} \right. \right. \nonumber \\
&& \left. \left. - \frac{ |e| E_{p} a_{0}}{E_{0}} (x+\xi(y)/a_{0} )  \right)  \right] \varphi^{y}_{n_{z\beta}}(x)=0 , 
 \label{eq14}
\end{eqnarray}
where $\infty >x \geq 0 $ and $\varphi^{y}_{n_{z\beta}}(0)=0$, as it is assumed that the wave function do not penetrate into LH.
We will look for a solution of Eq. (\ref{eq14}) using an expansion over the complete set of functions $\chi_{n}(x)$, within the interval $\infty >x \geq 0$, 
as
\begin{equation}
\varphi^{y}_{n_{z\beta}}(x)= \sum_{n=1}^{K} C^{(n)}_{n_{z\beta}}(y) \; \chi_{n}(x) , 
 \label{eq15}
\end{equation}
where a positive integer $K \to \infty$, and $\chi_{n}(x)$ satisfies  the equation for radial wave functions of the hydrogen atom 
with zero orbital quantum number \cite{LL59}
\begin{equation}
 \frac{d^{2}}{d x^{2}} \chi_{n}(x)= \left[ \frac{1}{n^{2}} - \frac{2}{x}  \right] \chi_{n}(x) .
 \label{eq16}
\end{equation}
Here
\begin{equation}
 \int_{0}^{\infty} dx \; \chi_{m}(x)  \chi_{n}(x) = \delta_{m,n}  \;, 
 \label{eq17}
\end{equation}
where $\delta_{m,n}$ is the Kronecker delta symbol, and \cite{LL59} 
\begin{equation}
 \chi_{n}(x) = \frac{2}{n^{5/2}} \; x \; e^{-x/n} L_{n-1}^{1}(\frac{2x}{n}) \; ,
 \label{eq18}
\end{equation}
where $L_{n-1}^{1}(2x/n)$ is the generalized polynomial of Laguerre \cite{gradshteyn1980}. 
In present study the generalized polynomial of Laguerre is defined as in  Ref. \cite{gradshteyn1980};
in Ref. \cite{LL59} this polynomial has somewhat different definition.

Point out that according to Eqs. (\ref{eq9}), (\ref{eq15}), (\ref{eq17})
\begin{equation}
a_{0}  \int_{0}^{\infty} dx \; [\varphi^{y}_{n_{z\beta}}(x)]^{2}= a_{0} \sum_{n=1}^{K} [C^{(n)}_{n_{z\beta}}(y)]^{2}=1 ,
 \label{eq19}
\end{equation}
in addition,  we also  will use dimensionless $\widetilde{\varphi}^{\;y}_{n_{z\beta}}(x)=a_{0}^{1/2}\varphi^{y}_{n_{z\beta}}(x)$
and $\widetilde{C}^{\;(n)}_{n_{z\beta}}(y)=a_{0}^{1/2} C^{(n)}_{n_{z\beta}}(y)$ that give
\begin{equation}
\int_{0}^{\infty} dx \; [\widetilde{\varphi}^{\; y}_{n_{z\beta}}(x)]^{2}= \sum_{n=1}^{K} [\widetilde{C}^{\; (n)}_{n_{z\beta}}(y)]^{2}=1 .
 \label{eq20}
\end{equation}
Assuming that $\mathcal{E}_{n_{z\beta}=1}(y)$ has a minimum value at $y=0$ and introducing 
a shorter notation $W_{n_{y\beta}}=\widetilde{W}_{n_{z\beta=1}, \;n_{y\beta}}$ from Eq. (\ref{eq11})
the conditions of a small non-adiabatic contribution are given  as
\begin{eqnarray}
&&(\mathcal{E}_{2}(0)-\mathcal{E}_{1}(0)) \gg (W_{2}-W_{1})  \gg 
\frac{\hbar^2}{2m_{0}}  \nonumber \\
&& \times | \int_{-\infty}^{\infty} dz \; \varphi_{n_{z\beta}=1}(z,y) \frac{\partial^{2}}{\partial y^{2}} \varphi_{n_{z\beta}=1}(z,y) | \;  .
 \label{eq21}
\end{eqnarray}
Eq. (\ref{eq21}) after using Eqs. (\ref{eq10}), (\ref{eq17}), (\ref{eq20}) obtains the form
\begin{eqnarray}
&&(\mathcal{E}_{2}(0)-\mathcal{E}_{1}(0)) \gg (W_{2}-W_{1})  \nonumber \\ 
&& \;\;\;\;\;\;\;\;\;\;\;\;\;\;\;\;\;\;\gg 
\frac{\hbar^2}{2m_{0}} \sum_{n=1}^{K} [\frac{d}{dy} \widetilde{C}^{\; (n)}_{n_{z\beta}=1}(y)]^{2} .
 \label{eq22}
\end{eqnarray}
Present study assumes that Eq. (\ref{eq22}) is satisfied. It shows that the non-adiabatic 
contributions are small and can be neglected.

Now using Eqs. (\ref{eq15}), (\ref{eq16}) in Eq. (\ref{eq14}) we have
\begin{eqnarray}
 &&\sum_{n=1}^{K} C^{(n)}_{n_{z\beta}}(y) \; \chi_{n}(x) \left[   
\frac{\mathcal{E}_{n_{z\beta}}(y)}{E_{0}}  +\frac{\Lambda_{1}/\Lambda}{x+d(y)/a_{0}}  \right. \nonumber \\
&& \left. +\frac{1}{2 n^{2}} - \frac{ |e| E_{p} a_{0}}{E_{0}} \left(x+\frac{\xi(y)}{a_{0}} \right)   \right] =0 \; . 
 \label{eq23}
\end{eqnarray}
Multiplying Eq. (\ref{eq23}) by $\chi_{m}(x)$, then integrating over $x$, $ \int_{0}^{\infty} dx$, and using Eq. (\ref{eq17}), we obtain
a system of $K$ linear homogeneous equations for $K$ unknown $C^{(n)}(y)$, for a given $y$, as follows 
\begin{eqnarray}
&&  C^{(m)}(y)  \left[   \frac{\mathcal{E}(y)}{E_{0}}  +\frac{1}{2 m^{2}} - \frac{ |e| E_{p} \xi(y)}{E_{0}} \right] \nonumber \\
&&+ \sum_{n=1}^{K} C^{(n)}(y)  \left[ <m| \frac{\Lambda_{1}/\Lambda}{x+d(y)/a_{0}}|n> \right. \nonumber \\
&& \left. -\frac{ |e| E_{p} a_{0}}{E_{0}} <m|x|n>   \right] =0 \; ,
 \label{eq24}
\end{eqnarray}
where a matrix element
\begin{equation}
<m| f(x) |n>= \int_{0}^{\infty} dx \; f(x) \; \chi_{m}(x)  \chi_{n}(x) . 
 \label{eq25}
\end{equation}

To solve a system of equations Eq. (\ref{eq24}), we assume a finite value $K_{0}$ for the positive integer $K$. 
Then Eq. (\ref{eq24}) presents a system $K_{0}$ linear homogeneous equations over $K_{0}$ unknown
$C^{(n)}(y)$ which will give nontrivial solution Eq. (\ref{eq15}) only if the determinant of Eq. (\ref{eq24}) is
nullified. The latter condition will give $K_{0}$ roots $\mathcal{E}_{n_{z\beta}}(y)$ that present energies of $K_{0}$ lowest
levels due to quantization along $z$, for a given $y$. For each such root $\mathcal{E}_{n_{z\beta}}(y)$ we have a set 
of $K_{0}$ amplitude functions $C^{(n)}_{n_{z\beta}}(y)$, with $n=1,...,K_{0}$.
If we, e.g., increase two times the value of $K_{0}$ then
the number of obtained energy levels also will increase two times as "older" levels will be obtained now with higher
precision. 

In particular, for $K_{0}=2$ assuming $d(y)/a_{0}=4$, $\varepsilon_{S}=5$ (then $\Lambda_{1}/\Lambda \approx 24.0$),
$E_{p}=0$ we obtain $<1|x|1>=1.5$,  $<2|x|2>=6.0$, $<1|x|2>=<2|x|1>= -0.55870$,
$<1|24/(x+4)|1>=4.4615$, $<2|24/(x+4)|2>=2.56528$, $<1|24/(x+4)|2>=<2|24/(x+4)|1>=0.335448$
it follows for the ground state level $\mathcal{E}_{1}(y)/E_{0}= - 5.010008$ (for a bulk LH  $\mathcal{E}_{1}/E_{0}= -0.5$)
and for the first exited level $\mathcal{E}_{2}(y)/E_{0}= - 2.64177$ (for a bulk LH  $\mathcal{E}_{2}/E_{0}= -0.125$).

\subsection{Self-consistent liquid helium film with a low-dimensional  electron system suspended 
on a substrate}

Now we present the rest of our self-consistent model. 
Following Ref. \cite{balev2005} we obtain that a profile, $\xi(y)$, of LH suspended on the substrate, $z=h(y)$, is defined 
by the nonlinear differential equation

\begin{eqnarray}
&&\!\!\! \negthickspace \!\!\!\! \frac{d^{2}\xi}{dy^{2}}-\left\{\frac{g\sigma}{\alpha_{ST}}[\xi(y)+H+\frac{|e| n(y) E_{p}^{\ast}(y)}{g\sigma}] \right.  \nonumber \\
&& \!\!\! \negthickspace \left. -\frac{\gamma}{\alpha_{ST}\left[\xi(y)-h(y)\right]^3}\right\}
\left[1+\left(\frac{d\xi}{dy}\right)^{2}\right]^{3/2} =0 ,
 \label{eq26}
\end{eqnarray}
where  $H>0$ defines the level of bulk LH as $z=-H$,
$\sigma=0.145 \; g/cm^{3}$ is the helium density, $\gamma=9.5 \times 10^{-15} \; erg$ is the vdW coupling constant
helium-substrate, $\alpha_{ST}=0.378 \; erg/cm^{2}$ is the surface tension of the liquid helium, and $g$ is the gravity acceleration.
Further,  we assume that only the fundamental quantum state $n_{z\beta}=1$ can be occupied, at any $y$, 
and that resulting 2DES (or 1DES) over LH is non-degenerate. Then 
the electronic density per unit area $n(y)$ obtains the form
\begin{eqnarray}
&&\!\!\!\!\!\!\!\!\!\!\!\!\! \negthickspace \!\!\!\! n(y)=\frac{2}{L_{x}}\sum_{k_{x\beta},\;n_{y\beta}}
e^{( \zeta -\widetilde{W}_{1, \;n_{y\beta}}-\frac{\hbar^{2}k_{x\beta}^{2}}{2m_{0}})/k_{B}T} |\Phi_{n_{y\beta}}(y;1)|^{2}   \nonumber \\
&&=\left(\frac{2 m_{0} k_{B} T}{\pi \hbar^{2}} \right)^{1/2} \sum_{n_{y\beta}}
e^{( \zeta-\widetilde{W}_{1, \;n_{y\beta}})/k_{B}T}  \nonumber \\
&&\;\;\; \times  |\Phi_{n_{y\beta}}(y;1)|^{2}  ,
 \label{eq27}
\end{eqnarray}
where $\zeta$ is the chemical potential,  the factor $2$  takes into account the spin degeneracy of the energy. 
Further, the effective electric field, $E_{p}^{\ast}(y)$, is given as
\begin{eqnarray}
E_{p}^{\ast}(y)&=&E_{p}+ \frac{\Lambda}{|e| a_{0}^{2}} \int_{0}^{\infty} \frac{dx}{x^{2}} [\widetilde{\varphi}_{1}^{\;y}(x)]^{2} \nonumber \\
&&+ \frac{\Lambda_{1}}{|e| a_{0}^{2}} \int_{0}^{\infty} \frac{dx}{\left[x+d(y)/a_{0}\right]^2} [\widetilde{\varphi}_{1}^{\; y}(x)]^{2} .
 \label{eq28}
\end{eqnarray}

The electron area density Eq. (\ref{eq27}) is obtained by integration of the bulk microscopic electron density 
$n_{bu}(y,z)$ over $z$ within a region of localization along $z$ of the probability distribution 
$\varphi^{2}_{1}(z,y)$ and taking into account of Eq.(\ref{eq9}). 
By integrating of Eq. (\ref{eq27}) over $y$, from $-\Delta L_{y}/2$ to $\Delta L_{y}/2$, and dividing the result by $\Delta L_{y}$ 
we have the average electron density within the main super cell as
\begin{equation}
\bar{n}= \frac{1}{ \hbar \Delta L_{y}} \left(\frac{2 m_{0} k_{B} T}{\pi} \right)^{1/2} \sum_{n_{y\beta}}
e^{( \zeta-\widetilde{W}_{1, \;n_{y\beta}})/k_{B}T} .
 \label{eq29}
\end{equation}
Then the total number of electrons  within the main super cell $N_{tot} = \bar{n} L_{x} \Delta L_{y}$ 
and the linear density within a super cell is given as
\begin{equation}
n_{L} =
\frac{1}{ \hbar } \left(\frac{2 m_{0} k_{B} T}{\pi} \right)^{1/2} \sum_{n_{y\beta}}
e^{( \zeta -\widetilde{W}_{1, \;n_{y\beta}})/k_{B}T} .
 \label{eq30}
\end{equation}

If in the main region potential Eq. (\ref{eq3}) is independent of $y$ then any $\mathcal{E}_{n_{z\beta}}$ also become
independent of $y$ and from Eqs. (\ref{eq7}), (\ref{eq12}) it follows that 
instead of $n_{y\beta}$ we can use the wave number 
$k_{y\beta}$. Then, e.g., for the fundamental energy level  
$\mathcal{E}_{1} <0$ we have that $\Phi_{k_{y\beta}}(y;1) =e^{ik_{y\beta}y}/\sqrt{L_{y}}$, and 
$\widetilde{W}_{1, \; k_{y}}=\mathcal{E}_{1} +\hbar^{2}k_{y\beta}^{2}/2m_{0}$. Then 
from Eq. (\ref{eq27}) it follows $n(y) = \bar{n}$ and
\begin{equation}
n = \frac{m_{0} k_{B} T}{\pi \hbar^{2}} e^{( \zeta+|\mathcal{E}_{1}|)/k_{B}T} ,
 \label{eq31}
\end{equation}
where $e^{( \zeta+|\mathcal{E}_{1}|)/k_{B}T} \ll 1$, as electrons are nondegenerate.
Point out these conditions correspond to $h_{1}=0$, for the present model of substrate, Eq. (\ref{eq1}).
Notice that here we have self-consistent Eqs. (\ref{eq15}), (\ref{eq24}), (\ref{eq26}), (\ref{eq28}), (\ref{eq31}) that
define $\mathcal{E}_{1}$ (along with relevant wave function, used in Eq. (\ref{eq28})), $\xi$, $E_{p}^{\ast}$, and
$\zeta$ para given $H$, $n$ (or $n_{L}$), $E_{p}$,  and $T$. It is seen that even for this rather simple problem
(e.g., Eq. (\ref{eq26}) reduces to an algebraic equation) 
%with second unknown $E_{p}^{\ast}$ that also depends on $\xi$)
the self-consistent set of Eqs. (\ref{eq15}), (\ref{eq24}), (\ref{eq26}), (\ref{eq28}), (\ref{eq31}) typically will not allow analytical solution.

It is seen that a self-consistent problem becomes a lot more complex if a substrate profile $h(y)$,  Eq. (\ref{eq1}), 
parameters, $h_{1}$ and $a$, are finite. Below we study this problem assuming a finite $\Delta L_{y}$ as well. For Eq. (\ref{eq26}) 
two boundary conditions, imposed at the boundaries of the main super cell, obtain the form 
\begin{equation}
d \xi(\pm \Delta L_{y}/2) /dy =  0 .
 \label{eq32}
\end{equation}

%%%************

%%%************

%%%************

\section{One-dimensional electron systems in self-consistent quantum wires
suspended on dielectric substrates: Results and discussion}

In below study it is assumed  that $T=0.6$K, 
%the dielectric constant of substrate 
$\varepsilon_{S}=5$, 
%the linear density within the main super cell 
$n_{L}=2\times 10^{3} cm^{-1}$, and 
%the external (holding) electric field 
$E_{p}=5 V/cm$.
These conditions are used in Figs. \ref{fig2} - \ref{fig11}. In addition, in Figs. \ref{fig2} - \ref{fig11} different substrate profiles $h(y)$, Eq. (\ref{eq1}),
are used; with different sets of a finite $h_{1}$, $a$, and $\Delta L_{y}$. Then we need to solve self-consistent 
 Eqs. (\ref{eq12}),(\ref{eq15}),(\ref{eq24}),(\ref{eq26})-(\ref{eq28}),(\ref{eq30}). Here, e.g., Eq. (\ref{eq12})  and 
Eq. (\ref{eq26}) are mutually coupled the second order nonlinear differential equations.  In this 
self-consistent coupling 
also  Eqs.(\ref{eq15}),(\ref{eq24}),(\ref{eq27})-(\ref{eq28}),(\ref{eq30}) are involved. For a given $H$
 we solve Eqs. (\ref{eq12}),(\ref{eq15}),(\ref{eq24}),(\ref{eq26})-(\ref{eq28}),(\ref{eq30})
by using a self-consistent numerical approach.  Notice, this problem can not be solved analytically. 

First, in Fig. \ref{fig2} it is assumed  the substrate profile $h(y)$ with $h_{1}=0.1 \mu$m, $a=1\mu$m, and $\Delta L_{y}=1 \mu$m.
We present: in Fig. \ref{fig2}(a) the self-consistent profile $\xi(y)$ of LH surface; in Fig. \ref{fig2}(b) the ground state energy $\mathcal{E}_{1}(y)$ of
transverse, mainly, quantization; in Fig. \ref{fig2}(c) the surface electron density $n(y)$, for obtained 1DES; in Fig. \ref{fig2}(d) the effective
electric  field $E_{p}^{\ast}(y)$ on an electron in the ground state $\mathcal{E}_{1}(y)$. Point out that in Fig. \ref{fig2} only
60 \% (i.e., $|y| \leq 0.3 \mu$m) of the main super cell are shown to present more clearly the main features.

In Fig. \ref{fig3} we present the energy spectra as functions of $H$, within the region  $10\;$cm$\geq H \geq 0.05\;$cm, for conditions of Fig. \ref{fig2}.
In Fig. \ref{fig3}(a) the dashed and the dot-dot-dashed curves show that the gap between two the lowest levels due to
mainly transverse quantization is large enough to warrant a two dimensional behavior of electrons even for $H=0.05$cm
as here $(\mathcal{E}_{2}(0)-\mathcal{E}_{1}(0))/k_{B} \approx 15.2$K$\gg T=0.6$K; for $H=0.5$cm we obtain
$(\mathcal{E}_{2}(0)-\mathcal{E}_{1}(0))/k_{B} \approx 38.6$K. 
In Fig. \ref{fig3}(b) the solid curve show that the gap between the lowest two levels due to
mainly lateral quantization is large enough to warrant a one dimensional behavior of electrons even for $H=0.05$cm
as here $(W_{2}-W_{1})/k_{B} \approx 1.71$K and $\exp(-(W_{2}-W_{1})/k_{B}T) \approx 5.8 \times 10^{-2}$; for $H=0.5$cm it follows
$(W_{2}-W_{1})/k_{B} \approx 3.60$K and $\exp(-(W_{2}-W_{1})/k_{B}T) \approx 2.5 \times 10^{-3}$.

In Fig. \ref{fig3}(b) the dashed and the dotted curves plot  $2(W_{1}-\mathcal{E}_{1}(0))$,
and $(W_{3}-W_{2})$. As they are very close to the solid curve, plotting $(W_{2}-W_{1})$,  it shows that for actual regions
of energy and $y$ the effective lateral potential $\mathcal{E}_{1}(y)$ can be well approximated by a parabolic potential.
For $H=2.5$cm the gap $(W_{2}-W_{1})/k_{B} \approx 5.12$K, i.e., further essentially increases. However, for
larger $H=10$cm the gap shows a decrease, $(W_{2}-W_{1})/k_{B} \approx 4.45$K. Point out that the results of Fig. \ref{fig3}
 are in a good agreement with Fig. \ref{fig2}. In particular, from Fig. \ref{fig2}(c) it is seen that a density profile, $n(y)$, peak
becomes higher and narrower as $H$ grows from $H=0.05$cm to $H=2.5$cm. However, for $H=10$cm the peak decreases and widens,
in comparison with the one for $H=2.5$cm.
This qualitatively is well explained by  the LH profiles behavior on Fig. \ref{fig2}(a). Indeed, for $H=10$cm the LH profile 
is essentially closer to the substrate, as $|y|$ grows from $0$, in a wider lateral region than for $H=2.5$cm.
It leads to more soft lateral confinement as it is seen also from Fig. \ref{fig2}(b).
Point out that for $H=0.05$cm, $0.5$cm, $2.5$cm and $10$cm 
Fig. \ref{fig2}(c) and Fig. \ref{fig2}(b) show that the peaks of Fig. \ref{fig2}(c) are well approximated by  the Gaussian, 
$n(y) \propto \ell_{y}^{-1} \times \exp(-y^{2}/\ell_{y}^{2})$, with  $\ell_{y} \lesssim 32$nm, $22$nm, $18$nm and $20$nm, respectively.  
In particular, it follows that the maximum density $n(0) \propto \ell_{y}^{-1}$
and $\ell_{y} \propto (W_{2}-W_{1})^{-1/2}$.
In addition, Figs. \ref{fig2}(b), \ref{fig2}(d) show that transverse confinement 
of 1DES becomes stronger as $H$ grows, and the effective electric field, $E^{*}_{p}(y)$, increases.

In Fig. \ref{fig4} we present the same dependences as in Fig. \ref{fig2} assuming the same parameters as in Fig. \ref{fig2} except of $\Delta L_{y}=10 \mu$m.
 Point out that in Fig. \ref{fig4} only a small part, $|y| \leq 0.3 \mu$m, of the main super cell, $|y| \leq 5\mu$m, is shown 
 to present more clearly the main features.
In Fig. \ref{fig5} we present the same energy spectra as in Fig. \ref{fig3} as functions of $H$, at  $10\;$cm$\geq H \geq 0.05\;$cm, for conditions of Fig. \ref{fig4}.
In Fig. \ref{fig5}(a) the dashed and the dot-dot-dashed curves show that the gap between the lowest two levels due to
mainly transverse quantization is large enough to warrant a two dimensional behavior of electrons.
Indeed,  for $H=0.05$cm it follows that $(\mathcal{E}_{2}(0)-\mathcal{E}_{1}(0))/k_{B} \approx 51.2$K and with an increase of $H$ the gap further
grows; it keeps close to $90$K for $H \geq 0.5$cm.

 In Fig. \ref{fig5}(b) the solid curve show that the gap between the lowest two levels due to
mainly lateral quantization is large enough to warrant a one dimensional behavior of electrons for $H=0.05$cm
as here $(W_{2}-W_{1})/k_{B} \approx 4.37$K and $\exp(-(W_{2}-W_{1})/k_{B}T) \approx 6.8 \times 10^{-4}$. For growing $H$ it follows
that $(W_{2}-W_{1})$ decreases.  However, it is still large enough to warrant a one dimensional behavior of electrons.
 In addition, in Fig. \ref{fig5}(b)  we observe that the solid, the dashed, and the dotted curves are very close. This shows that for actual regions
of energy and $y$ the effective lateral potential $\mathcal{E}_{1}(y)$ can be well approximated by a parabolic potential. 
  In agreement with Fig. \ref{fig5}(b),  in Fig. \ref{fig4}(c) a density profile, $n(y)$, peak becomes lower and wider as $H$ grows from $H=0.05$cm to
  $H=10$cm. This qualitatively is well explained by  the LH profiles behavior on Fig. \ref{fig4}(a). Indeed, due to larger $\Delta L_{y}$ in Fig. \ref{fig4} than 
  in Fig. \ref{fig2} a LH profile is essentially closer to the substrate within an actual lateral region, as $|y|$ grows from $0$, and this region
  is essentially wider, for the same $H$,  than in Fig. \ref{fig2}(a). Point out that qualitatively similar dependence on $H$ holds in Figs. \ref{fig2}, \ref{fig3}
  only as $H$ grows from $H=2.5$cm to $H=10$cm. The peaks of Fig. \ref{fig4}(c) are well approximated by  
 the Gaussian, $n(y) \propto \ell_{y}^{-1} \times \exp(-y^{2}/\ell_{y}^{2})$, with $\ell_{y} \approx 20$nm, $21$nm, $25$nm and $28$nm for $H=0.05$cm, $0.5$cm, $2.5$cm and $10$cm, respectively.

In Fig. \ref{fig6} we present the same dependences as in Fig. \ref{fig2} assuming the same parameters as in Fig. \ref{fig2} apart from $h_{1}=0.25 \mu$m.
 Point out that in Fig. \ref{fig6} only a part, $|y| \leq 0.3 \mu$m, of the main super cell, $|y| \leq 0.5\mu$m, is shown to present more clearly the main features.
In Fig. \ref{fig7} we present the same energy spectra as in Fig. \ref{fig3} as functions of $H$, at  $10\;$cm$\geq H \geq 0.05\;$cm, for conditions of Fig. \ref{fig6}.
Figs. \ref{fig6}, \ref{fig7} show qualitatively the same behavior as Figs. \ref{fig2}, \ref{fig3}.  In particular,  Fig. \ref{fig7}(b) shows the same qualitative
 behavior as Fig. \ref{fig3}(b). Increase in the height of modulation of the substrate from  $h_{1}=0.1\mu$m to $h_{1}=0.25\mu$m 
 leads to higher electronic concentration in the center of the channel, $n(0)$. Then, e.g., considering $H=10$cm, the minimum LH 
 film thickness, $d(0)=\xi(0)-h_{1}$, decreases from $9.4$nm, for $h_{1}=0.1\mu$m, to $7.5$nm, for $h_{1}=0.25\mu$m.

Further, in Fig. \ref{fig7}(a) the dashed and the dot-dot-dashed curves show that the gap between two the lowest levels due to
mainly transverse quantization is large enough to warrant a two dimensional behavior of electrons.
Indeed,  for $H=0.05$cm and  $0.5$cm it follows that $(\mathcal{E}_{2}(0)-\mathcal{E}_{1}(0))/k_{B} \approx 19.5$K and $47.5$K. With an increase of $H$ the gap further
grows. In Fig. \ref{fig7}(b) the solid curve show that the gap between the lowest two levels due to
mainly lateral quantization is large enough to warrant a one dimensional behavior of electrons even for $H=0.05$cm
as here $(W_{2}-W_{1})/k_{B} \approx 3.29$K and $\exp(-(W_{2}-W_{1})/k_{B}T) \approx 4.15 \times 10^{-3}$. For $H=0.5$cm it follows that
$(W_{2}-W_{1})/k_{B} \approx 6.55$K and $\exp(-(W_{2}-W_{1})/k_{B}T) \approx 1.8 \times 10^{-5}$.
 Point out that in Fig. 6(c) a density profile peak is well approximated as the Gaussian, 
 $n(y) \propto \ell_{y}^{-1} \times \exp(-y^{2}/\ell_{y}^{2})$, with $\ell_{y}$ decreasing 
 from $23$nm to $16$nm, and $14$nm as $H$ grows from $0.05$cm to $0.5$cm, and $2.5$cm.

In Fig. \ref{fig8} we present the same dependences as in Fig. \ref{fig6} assuming the same parameters as in Fig. \ref{fig6} except for $\Delta L_{y}=10 \mu$m.
 Point out that in Fig. \ref{fig8} only a small part, $|y| \leq 0.3 \mu$m, of the main super cell, $|y| \leq 5\mu$m, is shown.
In Fig. \ref{fig9} we present the same energy spectra as in Fig. \ref{fig7} as functions of $H$, at  $10\;$cm$\geq H \geq 0.05\;$cm, for conditions of Fig. \ref{fig8}.
Point out that Figs. \ref{fig8}, \ref{fig9} show qualitatively the same behavior as Figs. \ref{fig4}, \ref{fig5}.
Increase in a height of substrate modulation from  $h_{1}=0.1\mu$m to $h_{1}=0.25\mu$m 
 leads to higher electronic concentration in the center of the channel, $n(0)$. Then, considering $H=10$cm, the minimum LH 
 film thickness, $d(0)$, decreases from $8.5$nm, for $h_{1}=0.1\mu$m, in Fig. \ref{fig4}(a) to $6.4$nm, for $h_{1}=0.25\mu$m, in Fig. \ref{fig8}(a). 

In Fig. \ref{fig9}(a) the dashed and the dot-dot-dashed curves show that the gap between the lowest two levels due to
mainly transverse quantization is large enough to warrant a two dimensional behavior of electrons.
Indeed,  for $H=0.05$cm it follows that $(\mathcal{E}_{2}(0)-\mathcal{E}_{1}(0))/k_{B} \approx 107$K and for $H \geq 0.5$cm the
gap increases a bit and becomes $\approx 120$K. 
 In Fig. \ref{fig9}(b) the solid curve shows that the gap between the lowest two levels due to
mainly lateral quantization is large enough to warrant a one dimensional behavior of electrons for $H=0.05$cm
as here $(W_{2}-W_{1})/k_{B} \approx 9.02$K and $\exp(-(W_{2}-W_{1})/k_{B}T) \approx 2.9 \times 10^{-7}$. In addition, 
even though $(W_{2}-W_{1})$ decreases for growing $H$ it is still large enough to warrant a one dimensional behavior of electrons.
In agreement with this,  in Fig. \ref{fig8}(c) a density profile peak is the narrowest and the highest for $ H = 0.05$cm, 
and it becomes wider and lower as $H$ grows. These peaks are well approximated by the  Gaussian, 
$n(y) \propto \ell_{y}^{-1} \times \exp(-y^{2}/\ell_{y}^{2})$, with $\ell_{y} \approx 13$nm for $H=0.05$cm 
and $\ell_{y }\approx 20$nm for $H \geq 0.5$cm.

In Figs. \ref{fig10} and \ref{fig11} we present the same dependences as in Figs. \ref{fig6}, \ref{fig7} assuming,  except for $h_{1}=0.025\mu$m, $a=0.1 \mu$m and $ 10$ cm  $\geq H \geq 0.5$ cm,
the same parameters. I.e., in comparison with Figs. \ref{fig6}, \ref{fig7}, in Figs. \ref{fig10}, \ref{fig11} the parameters $h_{1}$, $a$ are reduced by ten times as 
the super cell size is kept constant,  at $\Delta L_{y}=1\mu$m.  Point out that in Fig. \ref{fig10} only a small part, $|y| \leq 0.1 \mu$m, of the main 
super cell, $|y| \leq 0.5 \mu$m, is shown.
Fig. \ref{fig10} shows qualitatively similar behavior with Fig. \ref{fig6} for $H=0.5$, $2.5$, and $10$cm.  
In particular, in Fig. 11(a) the dashed and the dot-dot-dashed curves show that the gap between two the lowest levels due to
mainly transverse quantization is large enough to warrant a two dimensional behavior of electrons.
Indeed,  for $H=0.5$cm it follows that $(\mathcal{E}_{2}(0)-\mathcal{E}_{1}(0))/k_{B} \approx 39.4$K; 
with further increase of $H$ the gap grows rapidly.

In Fig. \ref{fig11}(b) the solid (dashed) curve shows that $W_{2}-W_{1}$ ($2(W_{1}-\mathcal{E}_{1}(0))$) 
for $H=0.5$cm, $2.5$cm, and $10$cm is given as 
$14.9$K ($17.2$K), $34.9$K ($39.2$K), and $39.2$K ($41.9$K).
Here the dotted curve shows that $W_{3}-W_{2}$ is given, respectively, as  $7.48$K, $26.9$K, and $33.3$K.
These shows that for $H \gtrsim 0.5$cm all essential conditions of our treatment are satisfied.
Fig. \ref{fig11}(b) shows that for $H \gtrsim 0.5$cm the gaps between three the lowest levels due to
mainly lateral quantization is large enough to warrant a one dimensional behavior of electrons. In particular, 
for $H=0.5$cm and $10$cm, we have $\exp(-(W_{2}-W_{1})/k_{B}T) \approx 2.7 \times 10^{-11}$ and $\approx 4.2 \times 10^{-29}$. 
Fig. \ref{fig10}(c) shows that the density profile $n(y)$ peak is well approximated by the Gaussian, 
$n(y) \propto \ell_{y}^{-1} \times \exp(-y^{2}/\ell_{y}^{2})$, 
with $\ell_{y}$ decreasing from $\approx 10.4$nm to $\approx 6.6$nm as $H$ grows from $0.5$cm to $10$cm.
Fig. \ref{fig10}(c) shows that the peak becomes narrower and higher as $H$ grows.

Point out that in present figures a maximum  electron density, $n(y=0)$, for the obtained 1DESs is not too high as these 1DESs
are non-degenerated. It is in agreement with the assumed conditions.   
Notice that for QWs studied in Figs. \ref{fig2} - \ref{fig11} effect of tunnel coupling between super cells
on obtained self-consistent 1DESs, localized at $y=0$, is negligible already for extremely week disorder effects
on pertinent energy levels.

\section{Conclusions} 

We obtained a strong self-consistent enhancement of the transverse and the lateral quantizations
of an electron on LH suspended over the specially modulated dielectric substrates. The
enhancement  is due to a strong mutual interplay between the transverse and the lateral
movements of an electron.  
It is related also with self-consistent dependences of LH profile and of LH thickness over the substrate.
%They contribute to a strong enhancement of the transverse and the lateral quantizations of an electron.
Strong enhancements for the effective electron image potential and electric field
are obtained  due to a self-consistent modification of LH thickness.

%In particular, an essential modification of the effective electron potential
%is obtained  from a strong change of the image potential, due to a self-consistent modification of the LH thickness.

Non-degenerated 1DESs are obtained, at relatively high temperature $T=0.6$K and rather weak an external electric field
$E_{p}=5$V/cm, for dielectric substrates with different nanoscale modulation.
%the characteristic scales, ($h_{1}$; $a/2$; $\Delta L_{y}$), of the substrate modulation, in particular, 
%such as: ($100$nm; $500$nm; $1 \mu$m), ($250$nm; $500$nm; $1 \mu$m), and ($25$nm; $100$nm; $1\mu$m).
In particular, for $H=0.5$cm  in studied model setups we obtained that the gaps between two the lowest electron states
due to the lateral confinement $(W_{2}-W_{1})/k_{B}$ appear within the interval from $3.60$K  to $14.9$K 
as the gaps between two the lowest electron levels due to the transverse confinement 
$(\mathcal{E}_{2}(0)-\mathcal{E}_{1}(0))/k_{B}$ appear within the interval from $38.6$K  to $120$K.

We demonstrated that in an actual region the effective lateral potential $\mathcal{E}_{1}(y)$ for a 1DES is 
very close to a parabolic one. In addition, the electron density is well approximated as
$n(y) \propto \ell_{y}^{-1} \times \exp(-y^{2}/\ell_{y}^{2})$, with $\ell_{y}$ given in the range from $6.6$nm to
$32$nm. This form of $n(y)$ is qualitatively similar with the probability to find an electron at the fundamental
Landau level \cite{LL59} $\propto \ell_{0}^{-1} \times \exp(-y^{2}/\ell_{0}^{2})$, where $\ell_{0}$ is the quantum magnetic length.
A strong  "long-range" effect of $\Delta L_{y}$ on the properties of a self-consistent 
1DES is shown  as $\Delta L_{y}$ decrease from $10 \mu$m to $1 \mu$m.

In addition, our study of present non-degenerated 1DESs, at $T=0.6$K, have shown that assumed linear density $n_{L}=2 \times 10^{3} cm^{-1}$ 
is relatively small. In a sense
that its influence on obtained LH profile, self-consistent transversal and lateral quantizations, a spatial form of normalized electron density, etc.
is very weak. Then we can speculate that present results also will well approximate results for $T<0.6$K. In particular, for $T \ll 1$K, 
i.e., for temperatures of particular interest for a quantum computer \cite{dykman1999,dykman2003}.

\section*{Acknowledgments}

%\begin{acknowledgments}
This work was supported by the Brazilian FAPEAM  (Funda\c{c}\~{a}o de Amparo \`{a} Pesquisa do Estado do
Amazonas) Grants: Universal Amazonas (Edital 021/2011), O. G. B..
%\end{acknowledgments}

\end{document}